\documentstyle[12pt]{article}

\newcommand{\DerLabel}[1]{}
\newcommand{\Kom}[1]{}

\def\Mittefrei#1#2{\hbox to \hsize{#1\hss#2}}
\def\thebibliography#1{\section*{References}
  \list
  {[\arabic{enumi}]}{\settowidth\labelwidth{[#1]}
  \leftmargin\labelwidth
  \advance\labelsep8pt
  \advance\leftmargin\labelsep\parsep0.0em\itemsep0.0em
  \usecounter{enumi}} \relax }
\renewcommand{\thesection}{\arabic{section}.}

\newcounter{punkt}
\newenvironment{Punkt}{\parskip0ex\parsep0.1ex
  \begin{list}{(\roman{punkt})}{\usecounter{punkt}
  \parsep0.1ex
  \labelwidth 1cm\labelsep 0.5cm\leftmargin1.8cm}}{\end{list}}
\newcounter{satz}[section]
\newenvironment{Satz}[2]{\pagebreak[3]
  \label{#2-x} {\refstepcounter{satz}\label{#2-y} }
  \vskip1ex \underbar{\bf \thesection\arabic{satz}. #1}
  \DerLabel{#2}\sl}{\vskip1ex}

\newcommand{\Absatz}[2]{\pagebreak[3]
  \label{#2-x} {\refstepcounter{satz}\label{#2-y} }
  \vskip1ex{\bf \thesection\arabic{satz}. #1}\DerLabel{#2}}
\newenvironment{Beweis}{\pagebreak[3]\jot0pt\abovedisplayskip0pt
 \abovedisplayshortskip0pt\belowdisplayskip0pt\belowdisplayshortskip0pt
 {\bf Proof: }}{\hfill $\Box$\vskip1ex}
\newenvironment{Matrix}[1]{\left( \begin{array}{#1}}{\end{array}\right)}
\newcommand{\refxy}[1]{\ref{#1-x}\ref{#1-y}\DerLabel{#1}}

\newcommand{\hyp}{\;\Phi_{\!\!\!\!\!\!\!\! 1\>\;\;1}}

\def\clsubset{\mathop{\kern0pt\subset}\limits _{\rm closed}  }
\def\densesubset{\mathop{\kern0pt\subset}\limits _{\rm dense}  }

\newcommand{\Domain}{{\rm D}}
\newcommand{\Dom}{{\cal D}}
\newcommand{\Dome}{{\cal E}}

\newcommand{\Span}{{\rm Span}}
\newcommand{\restricted}[1]{{_{\textstyle |#1}}}

\newcommand{\specm}[2]{ {\bf P}^{#1}_{#2} }
\newcommand{\tildeE}{ \tilde{\bf P} }
\newcommand{\cequ}{ {\rm C}_q }

\newcommand{\mult}{{\rm m}}

\newcommand{\RR}{{I\!\!R}}
\newcommand{\NN}{{I\!\!N}}
\newcommand{\ZZ}{{Z\!\!\!Z}}

\newcommand{\CC}{{\>l\!\!\!C}}

\newcommand{\A}{{\cal A}}     
\newcommand{\Pol}{{\cal A}}     
\newcommand{\AC}{{\bf A}}     

\newcommand{\DDR}{{\cal D}_{\RR_q^3}^X}
\newcommand{\QG}{\Pol_{SO_q(3)}}

\newcommand{\QGC}{{{\bf A}_{SO_q(3)}}}

\newcommand{\QGCU}{{\bf A}_{SU_{\sqrt q}(2)}}
\newcommand{\QS}{\Pol_{\RR_q^3}}
\newcommand{\UU}{{\cal U}}

\newcommand{\UUU}{\UU_{SO_q(3)}}

\newcommand{\Obs}{{\cal O}}
\newcommand{\Hilb}{{\cal H}}

\newcommand{\Rhat}[4]{\,{{\hat R}{}^{#1#2}}_{#3#4}\,}
\newcommand{\Rinv}[4]{\,{{\hat R^{-1}}{}^{#1#2}}_{#3#4}\,}

\newcommand{\M}[2]{{M^#1}_#2}

\newcommand{\einsq}{q-1+\frac 1q}
\newcommand{\zweiq}{q+\frac 1q}
\newcommand{\dreiq}{q+1+\frac 1q}

\newcommand{\e}{{\rm e}}
\newcommand{\k}{{\rm S}}
\newcommand{\bder}{\bar\partial}
\newcommand{\id}{{\rm id}}

\newcommand{\Def}[1]{{\it #1}}

\newcommand{\scalar}[2]{#1\!\cdot\! #2}
\newcommand{\cross}[3]{[#1\!{\times}\! #2]_#3}

\newcommand{\ket}[1]{\left| #1\right\rangle}
\newcommand{\bra}[1]{\left\langle #1\right|}
\newcommand{\bracket}[2]{\left\langle #1 | #2 \right\rangle}

\begin{document}
\thispagestyle{empty}

\Mittefrei{March 1994}{LMU-TPW 1994-5}
\vspace{\fill}


{\LARGE\bf\sc
\begin{center}
The Hilbert Space Representations for $SO_q(3)$-symmetric Quantum Mechanics
\end{center}}
\vspace{\fill}

\begin{center}
{\large\it  Wolfgang~Weich }\\[10pt]
{\small\it
Sektion Physik der Universit\"at M\"unchen, \\
Theresienstra\ss e 37, D-80333 M\"unchen, Germany }
\end{center}
\vspace{\fill}
\vspace{\fill}

\noindent {\bf Abstract: }
{\rm
The observable algebra $\Obs$ of $SO_q(3)$-symmetric
quantum mechanics is generated by the coordinates $P_i$
and $X_i$ of momentum and position spaces
(which are both isomorphic to the
$SO_q(3)$-covariant real quantum space $\RR_q^3$). Their
interrelations are determined with the quantum group covariant
differential calculus. For a quantum mechanical representation
of $\Obs$ on a Hilbert space essential self-adjointness of
specified observables and compatibility of the covariance of the
observable algebra with the action of the unitary continuous
corepresentation operator of the compact quantum matrix group
$SO_q(3)$ are required.
It is shown that each
such quantum mechanical representation extends uniquely to a self-adjoint
representation of $\Obs$. All these self-adjoint representations are
constructed.
As an example an $SO_q(3)$-invariant Coulomb potential is introduced,
the corresponding Hamiltonian proved to be essentially self-adjoint and
its negative eigenvalues calculated with the help of a $q$-deformed
Lenz vector.

}
\pagebreak


\section{Introduction}\label{Intro0}
Symmetries play an important role in physics.
Quantum groups as symmetries of physics had been discovered in
the context of integrable systems.
They are however expected to play a more general role \cite{MS}
and are investigated as symmetries of some non-commutative spacetime
\cite{WZ1,Weich,CSW}.
It is hoped that there will be a quantum group symmetric
formulation of four-dimensional quantum field theory.
On the way to quantum field theory with $q$-deformed symmetry
first one-particle quantum mechanics has to be understood.
In this paper the quantum mechanics of one particle, moving in
some $q$-deformed $\RR_q(3)$-space with the configuration
space symmetry $SO_q(3)$, is formulated and investigated.

The outline of this paper is as follows.

In section 2 
the compact matrix quantum group $SO_q(3)$ is
introduced.
To fix the notation also the notions of quantum group covariance,
corepresentation and covariant algebras are recalled. The
quantum space $\RR_q^3$ is described by the algebra
generated by its three coordinates $X_i$.

In the next section the observable algebra is introduced as
covariant differential operators acting on the functions over the
quantum space $\RR_q^3$. While $X_i$ act by multiplication the
momenta $P_i$ are constructed from the $SO_q(3)$-covariant
partial derivative operators and shown to describe also a
real quantum space $\RR_q^3$ which is henceforth understood as
momentum space. The relations between position and momentum operators
involve a scaling operator $\mu$ and the quantized universal enveloping
algebra dual to $SO_q(3)$ represented as angular momentum algebra.
Its operators can be characterized as those commuting with
both scalar observables $X^2$ and $P^2$.
The observable algebra exhibits a manifest symmetry between
position and momentum coordinates.

The observable algebra has to be represented on a Hilbert space. And the
symmetry has to be given by a unitary operator. It is also required
that the topological structure of the compact quantum matrix group
$SO_q(3)$ shows up in this representation. Therefore some facts about
unitary continuous corepresentations of compact quantum matrix
groups are recalled in this section \ref{SHil}

In section 5 
some basic notions on Hilbert space
representations of unbounded operator algebras are recalled. Then
the quantum mechanical representation of $\Obs$ is
defined and motivated. This definition guarantees the compatibility between
the topological structure of the compact matrix quantum group and
the covariance of the observable algebra. The scalar operators
$P^2$ and $X^2$ are required to be observable, i.~e. represented
by essentially self-adjoint operators.
In the rest of this section the consequences of this definition
are studied. It comes out that quantum mechanical representation
can always be uniquely extended to a self-adjoint representation.
And all these self-adjoint representations can be constructed.

Finally in section 6 
a basic result on
the self-adjointness of operators of the form $H=P^2+V$
--- to be understood as Hamilton operators of $SO_q(3)$-quantum
mechanics --- is derived and applied as $SO_q(3)$-symmetric
Coulomb potential problem.

\section{The $SO_q(3)$-symmetry}\label{Intro}
Geometrical manifolds can be characterized by algebras of complex
functions over these manifolds which are (in usual geometry)
commutative.
Quantum space algebras are obtained as deformations of polynomial
algebras over manifolds where the parameter $q$ governs the
non-commutativity of the quantum space algebras.
The reality of the manifold leads to a $*$-algebra structure for
the quantum space algebra.
Specializing the manifolds to algebraic Lie groups the
polynomial algebra generated by the Matrix entries of the
fundamental representation are $q$-deformed. The group structure
becomes encoded in an additional Hopf algebra structure
realized on these polynomials. In this way the quantum groups
can be defined. This will be reviewed here for the Lie group
$SO(3)$ which has a standard $q$-deformation \cite{Drin,FRT}.
Considering $q$-deformations of polynomials of representation
spaces of Lie groups one obtains comodule algebras on which
the quantum group algebras coact.

For the definition of the quantum group $SO_q(3)$ one
chooses usually the irreducible representation of $SO(3)$
generated by the orthonormal vectors
$\frac 1{\sqrt 2} (\vec e_x + i\vec e_y)$, $\vec e_z$ and
$\frac 1{\sqrt 2} (\vec e_x - i\vec e_y)$ in $\CC^3$.

The definition of the quantum group $SO_q(3)$ is based on
the fundamental invariant tensors which are $q$-deformations of
the corresponding invariant tensors of the Lie group $SO(3)$.

\Absatz{Fundamental tensors of $SO_q(3)$.}{fundamental tensors}
The deformation parameter $q$ is restricted by $q\ge 1$.
The $q$-antisymmetric tensor $\{\epsilon_{ijk}\}_{ijk=1,2,3}$
and the metric tensor $\{\gamma_{ij}\}_{ij=1,2,3}$
are defined by
\begin{eqnarray*}
 \epsilon_{1jk} = \begin{Matrix}{ccc}
               0 & 0 & 0 \\
               0 & 1 & \frac1q   \\
               0 & -1 & 0
        \end{Matrix}_{jk}, &
 &\epsilon_{2jk} = \begin{Matrix}{ccc}
          0 & 0 & -1 \\
          0 & {\scriptstyle \frac1{\sqrt q}-\sqrt q }& 0 \\
          1 & 0 & 0
        \end{Matrix}_{jk} ,\nonumber\\
 \epsilon_{3jk} = \begin{Matrix}{ccc}
           0 & 1 & 0 \\
          -q & 0 & 0 \\
           0 & 0 & 0
        \end{Matrix}_{jk}\> &\quad\mbox{ and }\quad
 &\gamma_{jk} = \begin{Matrix}{ccc}
           0 & 0 & \frac1{\sqrt q} \\
           0 & 1 & 0 \\
     \sqrt q & 0 & 0          \end{Matrix}_{jk} .
\end{eqnarray*}
Further the tensors with upper indices $\{\gamma^{ij}\}_{ij=1,2,3}$ and
$\{\epsilon^{ijk}\}_{ijk=1,2,3}$ are defined by
\[
 \gamma^{ij} := \gamma_{ij}\quad \mbox{and} \quad
 \epsilon^{ijk} := \epsilon_{ijk}.
\]

For $q=1$ the above defined tensors coincide with their
usual counterparts and are $q$-deformations of the
completely antisymmetric tensor (however multiplied by $i$)
and the canonical metric on $\CC^3$
(in the above chosen basis).

If not otherwise stated the sum convention for repeated indices (one lower,
one upper) is always implied. It can be easily checked that
$ \epsilon^{ijk} = \epsilon_{abc}\, \gamma^{ci}\gamma^{bj}\gamma^{ak} $
and $ \epsilon_{ibc}\, \gamma^{bc} = \epsilon_{abi}\, \gamma^{ab} = 0 $
hold.
The obvious abbreviations ${\epsilon_n}^{ab}:=\gamma_{nm}\epsilon^{mab}=
\epsilon_{nmp}\gamma^{pa}\gamma_{mb}$ and
${\epsilon_{ab}}^k:=\epsilon_{abl}\gamma^{lk}$ are introduced.
They fulfill ${\epsilon_n}^{ab}{\epsilon_{ab}}^m = (\zweiq) \delta^m_n$ and
$\epsilon^{lkn}{\epsilon_n}^{ab}{\epsilon_{ka}}^m = (\einsq) \epsilon^{lmb}$.

These equations imply that the matrices
$(\frac 1\zweiq\,\epsilon^{ija}\epsilon_{akl})$ and
$(\frac 1\dreiq\,\gamma^{ij}\gamma_{kl})$
are orthogonal projectors on $\CC\otimes\CC$ with three- and
one-dimensional ranges respectively. They are understood as the
$q$-deformed antisymmetrizer and metric projector.

The $\hat R$-matrix
\[
 \Rhat ijkl := q\delta^i_k \delta^j_l - \epsilon^{ija}\epsilon_{akl}
 + (\frac 1q-1)\gamma^{ij}\gamma_{kl}
\]
with inverse
\[
 \Rinv ijkl = \frac 1q\delta^i_k \delta^j_l - \epsilon^{ija}\epsilon_{akl}
 + (q-1)\gamma^{ij}\gamma_{kl}
\]
obeys the Yang-Baxter-equation
$\Rhat ijab \Rhat bkcn \Rhat aclm = \Rhat jkde \Rhat idlf \Rhat femn$ which
states that this $\hat R$-matrix provides a representation of the
braid group on tensor products of $\CC^3$ \cite{Drin,FRT} and should
be understood as the deformed permutation action on these tensor products.

This preparation suggests the definition
of a $q$-deformation of the polynomial algebra over the Lie group $SO(3)$:

\begin{Satz}{Definition:}{qgroup}
The (unital, associative)
$\CC$-algebra $\QG$ is generated by the matrix elements
$\{\M ij\}_{i,j=1,2,3}$ of the Matrix $M$ with relations
\begin{eqnarray*}
 && \epsilon_{ijk}\, \M ia \M jb \M kc = \epsilon_{abc},\nonumber\\
 && \gamma_{ij}\, \M ia \M jb = \gamma_{ab},\\
 && \M ia \M jb\, \gamma^{ab} = \gamma^{ij}.\nonumber
\end{eqnarray*}
On these generators the $\CC$-antilinear and antimultilicative
involution $*$ is defined by
\[
 \left(\M ij \right)^* := \gamma^{jb}\, \M ab\, \gamma_{ai}
\]
making $\QG$ a $*$-algebra.
The tensor product $\QG\otimes\QG$ becomes also a $*$-algebra
with $(a\otimes b)(c\otimes d)=ac\otimes bd$ and
$(a\otimes b)^* = a^*\otimes b^*$ and the unit $1\otimes 1$.
On $\QG$ one defines the comultiplication
$\Delta:\, \QG\to\QG\otimes\QG$
and the counit $\e:\, \QG\to\CC$ as $*$-algebra homomorphisms
by
\begin{eqnarray*}
 &&\Delta(\M ij) = \M ik\otimes \M kj\nonumber\\
 &&\e(\M ij) = \delta^i_j
\end{eqnarray*}
and the $\CC$-linear antimultiplicative
antipode $\k:\, \QG\to\QG$ by
\[
 \k(\M ij)=(\M ji)^*.
\]
Thus $\QG$ becomes a $*$-Hopf algebra \cite{Abe},
the \Def{Hopf algebra $\QG$ of the quantum group $SO_q(3)$}.
\end{Satz}

For $q=1$, in the case of the undeformed group, one finds immediately
that in this definition $\A_{SO_1(3)}$
coincides with the polynomials of the matrix elements
of the fundamental representation of $SO(3)$.
For the Hopf algebra mappings one
gets for $g,h\in SO(3)$ and a polynomial $F$:
$\Delta(F)(g\otimes h)=F(gh)$, $\e(F)=F(\e)$ and $\k(F)(g)=F(g^{-1})$.
In the case $q=1$ one can also see that all the structure of the
algebraic group $SO(3)$ is completely contained in the $*$-Hopf algebra
$\QG$ as defined in \refxy{qgroup}.

\Absatz{The compact matrix quantum group $(\QGC,M)$.}{CMQGa}
There exists a $C^*$-algebra $\QGC$ in which the $*$-algebra
$\QG$ is densely imbedded and
the comultiplication $\Delta$ can be extended on $\QGC$ as a
$C^*$-homomorphism. Then $(\QGC,M)$ is a compact matrix quantum
group \cite{Wor}.

\begin{Beweis}
$\QG$ is the sub-$*$-Hopf algebra of even degree polynomials
of the $*$-Hopf algebra of $SU_{\sqrt q}(2)$ which has been proved
to belong to a compact matrix quantum group \cite{WorTw,Wor}.
\end{Beweis}

\Absatz{Corepresentations and covariant algebras.}{fin.corr}
The usual notion of representation can now be formulated for quantum
groups. For a Hopf algebra $\A$ an $\A$-corepresentation $(V,\Delta_R)$
consists of a $\CC$-vector space $V$ with a coaction
$\Delta_R:\, V\to V\otimes\A$ obeying
$(\Delta_R\otimes \id)\circ \Delta_R = (\id\otimes\Delta)\circ \Delta_R$
(coassociativity) and $(\id\otimes\e)\circ\Delta_R = \id$ (counit).
If $V$ is a Hilbert space with $v^\dagger:=\bracket{v}{.}$
compatible with the coaction, $({}^\dagger\otimes{}^*)\circ \Delta_R =
\Delta_R\circ{}^\dagger$, then $(V,\Delta_R)$ is called unitary.
$W\subset V$ is an $\A$-invariant subspace
if the restriction of $\Delta_R$
to $W$ defines an $\A$-corepresentation $(W,{\Delta_R}_{|W})$.
The $\A$-corepresentation $(V,\Delta_R)$ is called $\A$-irreducible
if the only $\A$-invariant subspaces are $V$ and $\{0\}$ \cite{Wor,Koor}.
If the $\A$-corepresentation space $V$ is an algebra and the coaction
$\Delta_R$ is an algebra homomorphism $(V,\Delta_R)$ is called an
$\A$-covariant algebra. If $V$ is in addition a $*$-algebra and
$({}^*\otimes{}^*)\circ \Delta_R = \Delta_R\circ{}^*$, then
$(V,\Delta_R)$ is called an \Def{$\A$-covariant $*$-algebra}.

For $\QG$ one has special corepresentations:
The \Def{scalar} corepresentation is given by a one-dimensional
vector space and the coaction $\Delta_R(.)=.\otimes 1$.
A vector corepresentation consists of a 3-dimensional
$\CC$-vector space $V$ with basis $\{A_i\}_{i=1,2,3}$ and the coaction
$\Delta_R(A_i)=A_j\otimes\M ji$ on them. This basis $\{A_i\}$ is called
an \Def{$SO_q(3)$-triplet}.
If the triplet $\{A_i\}$ is contained in an
$\QG$-covariant $*$-algebra and fulfills $A_i^*=\gamma^{ik}A_k$
then $\{A_i\}$ is called a \Def{real triplet}.
The real linear combinations will also be used later, they are
denoted $X_i {R^i}_\alpha =: X^R_\alpha$ for $\alpha=1,2,3$
with the matrix
\[\label{real basis}
 {R^i}_\alpha := \begin{Matrix}{ccc}
   \sqrt{\frac{q+1}{2(q+\frac 1q)}} & 0 & \sqrt{\frac{\frac1q+1}{2(q+\frac
1q)}}\\
           0 & 1 & 0 \\
     i\sqrt{\frac{q+1}{2(q+\frac 1q)}} &
                  0 & -i\sqrt{\frac{\frac1q+1}{2(q+\frac 1q)}}
          \end{Matrix}_{i\alpha}.
\]
Two (real) $SO_q(3)$-triplets contained in a
$\QG$-covariant ($*$-)algebra give rise
to the (real) $SO_q(3)$-scalar
$ \scalar AB := A_i B_j \gamma^{ij} $ and the (real) $SO_q(3)$-triplet
$ \cross ABk :=  {\epsilon_k}^{ij}\, A_i B_j$.


Having now defined the symmetry and its corepresentation one needs still
the notion of quantum space, the space on which these symmetries act. It will
be constructed from the fundamental corepresentation of the quantum group.
Again it is described by the polynomials of its coordinates.

\Absatz{The quantum space algebra $\QS^X$.}{qspace}
Let $\{X_i\}$ be a real triplet. Then the $\QG$-covariant $*$-algebra
$\QS^X$ is generated by $\{X_i\}_{i=1,2,3}$ with relations
\begin{equation}\label{XX}
 \cross XXk=0\qquad\mbox{and}\qquad
 X_i^*=\gamma^{ij}X_j.
\end{equation}

In $\QS^X$ the scalar $X^2:=\scalar XX$ is central and real,
i.~e. $(X^2)^*=X^2$. In the real basis it is
$X^2=\frac{q+\frac1q}2\,\sum_{\alpha=1}^3 (X^R_\alpha)^2$.


\section{Differential calculus and Observable algebra}\label{DC}
In non-commutative geometry differential calculus is defined
purely algebraically. For a covariant differential calculus
on a quantum space \cite{PuW,WZ1} one requires the existence of
covariant operators as deformations of partial derivatives.
For the quantum space $\RR_q^3$ this had been
worked out in \cite{CSW}.

\Absatz{Partial derivatives and scaling operator.}{partials}
The \Def{partial derivatives} \linebreak
$\{\bder_i\}_{i=1,2,3}$,
linear operators acting on $\QS^X$, are defined by
$\bder_i(1) = 0$ and the generalized Leibniz rule
$\bder_i(X_j\, f) =
   \gamma_{ij}\, f + X_a\bder_b \,\left(\frac 1q\Rhat abij\,f\right)$
for each $f\in\QS^X$.
The \Def{scaling operator}
$\mu:\, \QS^X\to\QS^X$ is the linear operator defined by
$\mu(1)=1$ and $\mu (X_i\,f)=q\,X_i\,\Lambda(f)$ for each $f\in\QS^X$.
The partial derivatives are covariant operators,
$\Delta_R\left(\bder_k\, (f)\right)=
 (\bder_m\otimes\M mk)\left(\Delta_R(f)\right)$,
the scaling operator acts as a scalar operator,
$\Delta_R\left(\mu(f)\right) =
 (\mu\otimes \id)\left(\Delta_R(f)\right)$.
The elements $X_i$,$\bder_j$ and $\mu$ act as linear operators on
$\QS^X$ by left multiplication and left action respectively.
They generate an algebra of operators on $\QS^X$ which
is denoted by $\DDR$.
The partial derivatives fulfill $\cross\bder\bder k=0$,
the relations with the scaling operator are given by
$\mu\bder_i=\frac1q \bder_i\mu$.
The scaling operator is invertible in $\DDR$ with
$ \mu^{-1}:=\mu\left(1+q^{-2}(1-q^{2}) \scalar X\bder +
   q^{-3}(1-q)^2 \scalar XX \,\scalar\bder\bder\right).$
Defining
$\partial_i :=
 \mu^2\left(\bder_i + (q^{-2}-q^{-1} X_i \,\scalar\bder\bder \right)$
one gets a second triplet $\{\partial_i\}$ of partial derivatives \cite{Ogi}.
It obeys $\partial_i\,X_j = \gamma_{ij} + X_a\,\partial_b \,q\Rinv abij.$
Introducing on $\DDR$ the $*$-involution defined on the generators by
$(X_i)^* = \gamma^{ij}\,X_j$,
$(\bder_i)^* = -q^3\gamma^{ij}\,\partial_j$ and
$\mu^* = q^{-3}\mu^{-1}$, the differential operator algebra
$\DDR$ becomes an $\QG$-covariant $*$-algebra \cite{CSW,OZ}.

\Absatz{The momentum coordinates.}{momenta}
In $\DDR$ one finds the real triplet $\{P_i\}$ with
\[
 P_i := \frac 1{i\, (1+q^{-3})} \left(\partial_i + q^{-3}\bder_i \right).
\]
which fulfills
\begin{equation}\label{PP}
 \cross PPk=0\qquad\mbox{and}\qquad
 P_i^*=\gamma^{ij}P_j
\end{equation}
and generates in $\DDR$ a covariant $*$-subalgebra
which is isomorphic to $\QS^X$.
(At this stage one finds only that the $P_i$ fulfill all the
defining relations for $\QS^X$. The isomorphy can be read off from
the symmetry \refxy{Aut} or the representations of this algebra
discussed later.) This new quantum space
will be understood as the quantum mechanical momentum space. This leads
to the definition of the observable algebra below in~\refxy{Obs}.

The relations between position and momentum coordinates, i.~e.
the $q$-Heisenberg relations, still have to be investigated:
\begin{eqnarray}\label{aa1}
 &i\left( P_a\, X_b - X_c\,P_d\,q\Rinv cdab \right)
     =: \mu^{-1}\left( \gamma_{ab}\, W + {\epsilon_{ab}}^m\,
        \frac{q-1}{q^2-q+1}\,L_m \right), & \nonumber\\
 &-i\left( X_a\, P_b - P_c\,X_d\,q\Rinv cdab \right)
     =  {\mu^*}^{-1}\left( \gamma_{ab}\, W + {\epsilon_{ab}}^m\,
        \frac{q-1}{q^2-q+1}\,L_m \right) &
\end{eqnarray}
with the real vector triplet $\{L_i\}$ and
the real scalar $W$ in $\DDR$.
The commutation relations with the coordinates and momenta are
\begin{eqnarray}
 & L_i\, Z_j = -{\epsilon_i}^{cd}{\epsilon_{dj}}^e \, Z_c\,L_e +
         {\epsilon_{ij}}^a\, Z_a\,W ,\qquad
 \scalar L Z = \scalar Z L = 0, &    \nonumber\\
 & W\,Z_j = ({\textstyle \einsq})\,Z_j\,W -
  {\textstyle (\sqrt q-\frac1{\sqrt q})}^2
   \,{\epsilon_j}^{rs}\,  Z_r\,L_s &
\end{eqnarray}
with $Z_i=X_i,P_i$. In particular it holds
$L_i Z^2 = Z^2 L_i$ and $W Z^2 = Z^2 W$. This suggests already the
name \Def{angular momentum algebra} for the algebra generated by $\{L_i,W\}$
with the relations
\begin{equation}
 \scalar LL = \frac{W^2-1}{(\sqrt q\!-\!\frac1{\sqrt q})^2},\qquad
 \cross LLk = L_k W,\qquad
 L_k W = W L_k
\end{equation}
and becomes justified by the following result \refxy{L1}.
The relations involving $\mu=q^{-3} {\mu^*}^{-1}$ are
\begin{equation} \label{aa4}
 \mu P_i = \frac 1q P_i\mu,\qquad
 \mu X_i = q X_i\mu,\qquad
 \mu L_i = L_i\mu,\qquad
 \mu W = W \mu.
\end{equation}

For the representation theory the following relations are crucial.

\Absatz{}{reconstr}
Considering the equation
$\cross X{[X\!\times\!\bder]}k = -X_k\,\scalar X\bder + \scalar XX\,\bder_k$
and the $*$-conjugated equation one derives in $\DDR$
\begin{equation} \label{reconstrP}
 X^2 P_i=\frac\cequ{i\,(q-1)}\left[
   q^{-1}X_iW\mu^{-1}-qW\mu^{-1}X_i + q^2X_iW\mu-W\mu X_i\right]
\end{equation}
resulting in
\begin{equation} \label{reconstrPP}
 X^2 P^2 =
  \cequ^2\,\left[ \frac{(1+q)^2}q\, W^2 - q\mu^2 - 2 - q^{-1}\mu^{-2} \right]
\end{equation}
with the constant $\cequ:=q^2\, (1+q^3)^{-1}\, (1-q)^{-1}$.

\Absatz{The dual algebra $\QG'$ of $\QG$.}{KV-alg}
The space $\QG'$ of linear functionals on $\QG$
becomes a $*$-algebra with multiplication
$\tilde\rho,\tilde\zeta\mapsto (\tilde\rho\otimes\tilde\zeta)\circ\Delta$,
unit $\tilde 1:=\e$
and involution
$\tilde\rho^*(.)=\overline{\tilde\rho\left(\left(S(.)\right)^*\right)}$.
The \Def{quantized universal enveloping algebra of $SO(3)$}.
\cite{Drin,Jim,FRT} is the $*$-subalgebra $\UUU$ of $\QG'$ generated by
$\{\tilde K^{\pm 1},\tilde S^\pm\}$.
These linear functionals on $\QG$ are defined
on the generators of $\QG$ by
\begin{eqnarray*}
 & \tilde K^{\pm 1}(\M ij) = \begin{Matrix}{ccc}
            q^{\mp 1} & 0 & 0 \\
               0 & 1 & 0   \\
               0 & 0 & q^{\pm 1}
        \end{Matrix}_{ij} ,&\\
 & \tilde L^+(\M ij) =  \begin{Matrix}{ccc}
           0 & \frac 1{\sqrt q}& 0 \\
          0 & 0 & -1 \\
          0 & 0 & 0
        \end{Matrix}_{ij} ,\qquad
 \tilde L^-(\M ij) = \begin{Matrix}{ccc}
           0 & 0 & 0 \\
          1 & 0 & 0 \\
          0 & -\sqrt q & 0
        \end{Matrix}_{ij} & \nonumber,
\end{eqnarray*}
while their action on elements of higher order
of $\QG$ is defined inductively by
\begin{eqnarray*}
 &&\tilde K^{\pm 1}(ab) = \tilde K^{\pm 1}(a)\, \tilde K^{\pm 1}(b),
 \nonumber\\
 && \tilde L^\pm(ab) = \e(a) \tilde L^\pm(b) + \tilde L^\pm(a) \tilde K(b).
\end{eqnarray*}
They fulfill
$\tilde L^\pm\, \tilde K = q^{\pm 1}\, \tilde K\, \tilde L^\pm$,
$\frac 1{\sqrt q} \tilde S^+\tilde S^- - \sqrt q \tilde S^-\tilde S^+
= \frac{\e-\tilde K^2}{q-\frac 1q}$,
$\tilde K^*=\tilde K$ and $(\tilde S^-)^* = \sqrt q\, \tilde S^+$.
This algebra possesses the real central element
\[
  \tilde W := \frac q{1+q}\, \tilde K^{-1} + \frac 1{1+q}\, \tilde K +
          q^{-\frac 12}(q\!-\!1)^2\,
          \tilde K^{-1} \,\tilde L^-\,\tilde L^+.
\]
Each (unitary) corepresentation $(V;\Delta_R)$ of $\QG$ is canonically
a ($*$-)representation of
$\QG'$ with $\QG'\ni\tilde\rho\mapsto (\id\otimes\tilde\rho)\circ\Delta_R$.

It is clear from this construction that operators $\tilde\rho$
thus lead always to operators on $\QG^X$ which commute with the action
of the scalar operators, in particular $X^2$ and $P^2$.
As examples one identifies as operators on $\QS^X$:
\[
 (W,L_1,L_2,L_3)=\left(\id\otimes
 \left(\tilde W,\tilde L^+,
 {\textstyle\frac1{\sqrt q-\frac1{\sqrt q}}}(\tilde K-\tilde W),
 \tilde L^-\right)\right)\circ\Delta_R
\]
preserving the $*$-structure.

The converse is also true, the property of commuting with $P^2$ and $X^2$
characterizes the elements of the symmetry algebra completely:

\begin{Satz}{Proposition.}{L1}
The algebra of linear operators on $\QS^X$ commuting with the action of the
operators $X^2,P^2\in\DDR$ is as an algebra isomorphic to
the $*$-algebra $\QG'$ of linear functionals on $\QG$:
For each operator $\rho:\, \QS^X\to\QS^X$ with $[\rho,X^2]=[\rho,P^2]=0$
there exists a unique $\tilde\rho\in\QG'$ such that
$\rho=(\id\otimes\tilde\rho)\circ\Delta_R$.
\end{Satz}

\begin{Beweis}
The vector space $\QS^X$ can be decomposed as
$\QS^X=\bigoplus_{n,l\in\NN_0,l} (X^2)^n\,Z_l$ where
$Z_l$ consists of homogeneous harmonic polynomials of degree $l$
in the kernel of $P^2$ \cite{HW}. From \refxy{reconstr} one reads
off that $Z_l$ is eigenspace of $W$ with eigenvalue
$c_l=\frac{q^{1+l}+q^{-l}}{q+1}$. Since the
homogeneous polynomials of $\QS^X$ carry a corepresentation of $\QG$ and
$X^2$ transforms trivially it follows that $Z_l$ is an invariant
space in $\QS^X$, therefore also a representation space of $\UUU$.
One proves at once that an irreducible representation of $\UUU$ with
$\tilde W\mapsto c_l$ is $2l+1$-dimensional like $Z_l$.
Choosing an orthonormal basis $\{\ket{lm}\}$ according to \refxy{lm-states}
this representation becomes a $*$-representation of $\UUU$.
That means that $(Z_l,{\Delta_R}_{|Z_L})$ becomes an irreducible unitary
corepresentation of $\QG$ with
$\Delta_R:\,\ket{lm}\mapsto \ket{lm'}\otimes t^l_{m'm}$ with
$t^l_{m'm}\in\QG$ homogeneous of degree $l$.
\cite{WorA} states that all the set $\{t^l_{m'm}\}_{l,m,m'}$
is linearly independent. From dimension counting one finds also that
this set spans already $\QG$.
$\{X^{2\alpha}\ket{lm}\}$ is a basis of $\QS^X$.
{}From the action of $X^2$ and $P^2$ (using the relation in \refxy{reconstr})
on this basis one derives
for $\rho$ with the properties required that
$\rho(X^{2\alpha} \ket{lm})=\sum_{m'}X^{2\alpha} \ket{lm'}R^l_{m'm}$
with some matrix coefficients $R^l_{m'm}$.
The linear operator
$\tilde\rho\in\QG'$: $t^l_{m'm}\mapsto R^l_{m'm}$ is then the solution.
\end{Beweis}

Summarizing the above the algebra of
observables of quantum mechanics on $\RR_q^3$
should be defined as follows.

\begin{Satz}{Definition:}{Obs}
The \Def{observable algebra} $\Obs$ of $SO_q(3)$-symmetric quantum
mechanics is defined to be
the $\QG$-covariant $*$-algebra which is
generated by the phase space coordinates
$X_i$ and $P_i$. \\
The \Def{position coordinates $X_i$} and the \Def{momentum coordinates
$P_i$} generate momentum and postition
quantum spaces isomorphic to $\QS^X$ (equations (\ref{XX},\ref{PP})).
Their $\QG$-covariant $q$-commutator
closes upto the scaling operator $\mu$ into the
angular momentum algebra, generated by the real triplet $\{L_i\}$
and the real scalar $W$, a subalgebra of the quantized
universal enveloping algebra $\UUU$, according to
equations (\ref{aa1}--\ref{aa4}).
\end{Satz}

\Absatz{A $*$-automorphism of $\Obs$.}{Aut}
The symmetry between momentum and position coordinates
appears as the $*$-algebra automorphism compatible with the
quantum group covariance defined by
\[
 P_i\mapsto -X_i,\qquad X_i\mapsto P_i
\]
which lets the angular momentum algebra invariant,
$S_i\mapsto S_i$, $W\mapsto W$, and maps the scaling operator on its
$*$-conjugate, $\mu\mapsto\mu^*$.

It should be noticed that $\mu^{\pm 2}\in\Obs$ but $\mu\not\in\Obs$.
Accordingly the operators $\mu^{\pm 1} W$, $\mu^{\pm 1} S_i$,
$W^2$, $WS_i$, $WK,...\in \Obs$ but $W,S_i\not\in\Obs$. However it will
turn out that in each representation in a Hilbert space to be considered
later a bounded Hilbert space operator $\mu$ can be uniquely
\Def{defined} which fulfills all the algebraic relations above.

\section{$\QG$-covariant Hilbert space}\label{SHil}
In quantum mechanics a symmetry is given by unitary transformations
on the Hilbert space of states and a covariant transformation law for
the observables \cite{MS}. A continuous symmetry is described
by a topological (more special a Lie) group $G$. This topological group is
then represented on the Hilbert space $H$ by a continuous mapping
$U:\,G\times H\to H$ which is also a group homomorphism.

Let us denote with $B(H)$ the bounded operators on the Hilbert space $H$,
let $C\!B(H)$ be the compact operators on $H$. Further let $M(A)$ be the
multiplier algebra of the $C^*$-algebra $A$.

\Absatz{Unitary continuous corepresentation.}{UCC}
A unitary continuous corepresentation $(U,H)$ of the compact quantum
matrix group $(\AC,u)$ on the Hilbert space $H$ is given by
a unitary operator
$U\in M(C\!B(H)\otimes \AC)\subset B(H\otimes\ell)$
(where $\ell$ is a Hilbert space on which $\AC$ is faithfully represented)
fulfilling the coaction property
$U_{12} U_{13} = (\id\otimes\Delta)(U).$
($U_{12}:=U\otimes 1$, $U_{13}:=(\id\otimes\sigma)(U_{12})$
with the linear
transposition operator $\sigma: A\otimes B\mapsto B\otimes A$.)
\cite{PW}.

For a usual compact group $G$ a unitary continuous representation
would be a continuous mapping $u: G\times H\to H$,
$g,v\mapsto u(g)v$ or $U: G\to B(H)$, $g\mapsto U(g)$,
i. e. an operator valued continuous (in the strong operator
topology) function on $G$, and one could prove
that this definition reproduces the above definition.

\begin{Satz}{Proposition:}{decomp.}
Let $(U,H)$ be a unitary continuous corepresentation of $(\AC,u)$.
Then it decomposes in
finite dimensional irreducible corepresentations $(U_\rho,K_\rho)$, i.~e.
$H=\bigoplus_{\rho\in I} K_\rho$ with finite dimensional
subspaces of $H$ for a certain index set $I$. If $E_\rho$ is the orthogonal
projector on $K_\rho$ then the restriction $U_\rho := U\,(E_\rho\otimes 1)$
belongs to $B(K_\rho)\otimes\A$ where $\A$ is the $*$-algebra generated
by the matrix entries of $u$.
\end{Satz}

\begin{Beweis}
This was proved first in \cite{PW}.
A derivation going along the proof of the classical result
for usual compact groups \cite{Knapp} has been given in
\cite{DW}.
\end{Beweis}

This can now be applied to the compact matrix quantum group $SO_q(3)$.

\Absatz{Hilbert space $*$-representations of $\UUU$.}{lm-states}
Let $(U,H)$ be a unitary continuous corepresentation of $(\QGC,M)$.
Let $(U_\rho,H_\rho)$ be as in \refxy{decomp.}. Then
one defines on the domain
$\Span(\cup_{\rho\in I} H_\rho)\densesubset H$
a $*$-representation of $\QG'$ by
$\tilde\chi\mapsto \sum_\rho(\id\otimes\tilde\chi)(U_\rho)$
by (possibly unbounded) closable operators on this domain.
Let their closures be denoted by
\[
 \tilde\chi^U := \overline{\sum_\rho(\id\otimes\tilde\chi)(U_\rho)}.
\]
The attention is restricted now to $\UUU$.
The operators $\tilde K^U$ and $\tilde W^U$ restricted to some $H_\rho$ are
self-adjoint and commuting, hence $\tilde K^U$ and $\tilde W^U$
are self-adjoint and simultanously diagonalizable
on the whole Hilbert space $H$.
{}From these $*$-representation properties one derives:
The Hilbert space decomposes in irreducible representations of $\UUU$
which are classified by $l\in\frac 12\NN_0$ and $\sigma=\pm 1$:
$H_{l,\sigma}=\Span\{\ket{l,m}| m=-l,-l+1,...,l\}$ is the finite-dimensional
Hilbert space with orthonormal basis vectors $\ket{l,m}$ \cite{Rosso},
\begin{eqnarray*}
&& \tilde K^U\ket{l,m}=\sigma q^{-l}\>\ket{l,m}, \nonumber\\
&& \tilde W^U\ket{l,m}= \sigma c_l\>\ket{l,m} \nonumber\\
&&\quad\mbox{with}\;c_l=\frac{q^{1+l}+q^{-l}}{q+1} =
  1 + [l\!+\!1]_{\sqrt q}[l]_{\sqrt q}
  {\textstyle(\sqrt q\!-\!\frac1{\sqrt q})^2} ,\\
&& (\tilde L^\pm)^U\ket{l,m}= \alpha_{l,m\pm 1}\,q^{\mp\frac12}\,
   \ket{l,m\pm 1}\nonumber \\
&&\quad\mbox{with}\; \alpha_{l,m\pm1} =
    \frac{q^{-\frac{m\pm1}2}}{[2]_{\sqrt q} } \,
    \sqrt{[l]_{\sqrt q}[l+1]_{\sqrt q} -
    [m]_{\sqrt q}[m\pm 1]_{\sqrt q} }\>.  \nonumber
\end{eqnarray*}
Here the \Def{symmetric $q$-numbers} had been introduced
\[
 [x]_p := \frac{p^x-p^{-x}}{p-p^{-1}} .
\]

After this preparation
the operator $U$ can be given explicitly. From the proof of \refxy{L1}
the inequivalent unitary corepresentations of $\QG$ are known.
They correspond to the integer values for $l$. Furthermore this
fixes $\sigma=+1$. Thus there exists an orthonormal Hilbert space basis of
$H$ consisting of simultanous eigenvectors of $\tilde W^U$ and $\tilde K^U$:
$\{\ket{\alpha,l,m}:\>\alpha\in A_l;|m|,l\in\NN_0;|m|\le l\}$ with
an $l$-dependent index set $A_l$. The unitary operator $U$ becomes
\begin{eqnarray*}
 U &=& \sum_{\alpha,l,m,m'}
 \ket{\alpha,l,m'}\bra{\alpha,l,m} \otimes t^l_{m'm}\nonumber\\
 &=& \sum_{l,m,m'}
  \prod_{\mu=0}^{l-m'}\left(\frac{(\tilde L^-)^U}{\alpha_{l,l-\mu}}\right)\,
   \tildeE^U_{ll} \,
  \prod_{\mu=0}^{l-m}
     \left(\frac{(\tilde L^-)^U}{\alpha_{l,l-\mu}}\right)^\dagger\,
   \otimes t^l_{m'm}
\end{eqnarray*}
with the orthogonal Hilbert space projector $\tildeE^U_{ll}$ projecting
on the simultanous \linebreak[4]
eigenspaces of $\tilde W^U$ and $\tilde K^U$ with
eigenvalues $c_l$ and $q^{-l}$.
The sum has to be understood in the strong operator topology. The
matrices $(t^l)_{m'm}$ belong to the $2l+1$-dimensional corepresentation
of $\QG$, the unitarity is reflected by $S(t^l_{mm'})={t^l_{m'm}}^*.$

\section{Hilbert space of $SO_q(3)$-symmetric quantum mechanics}
\label{HilbQM}
To formulate the definition of $SO_q(3)$-quantum mechanics
some further points need to be mentioned.
The operators in $\Obs$ are like in the undeformed case ($q=1$)
not bounded. Since they are to be represented on a Hilbert space
their domain can not be the full Hilbert space. Instead a domain is
to be defined for each operator, and symmetric operators
(i.~e. those operators $O$ which coincide with their
Hilbert space adjoints restricted to the domain of $O$) need not
be diagonalizable, i.~e. have a projector valued spectral measure.

On the other hand the most fundamental property of an observable
is its measurability. That means that a measurement projects
any state vector on an eigenspace (or approximate eigenspace in the
case of an observable with continuous spectrum) of
the measured observable, and the spectrum is real.
This property of diagonalizability is fulfilled by self-adjoint
operators, i.~e. those which coincide with their Hilbert space adjoints.

To represent the observable algebra
there must be a common dense domain
$\Domain$ of all the operators where the
algebra relations of $\Obs$ are fulfilled \cite{Schmud},
the involution $*$ becomes represented by the Hilbert space adjoint.
The observables are essentially self-adjoint on this domain.

\Absatz{Hilbert space representations.}{notation1}
Let $\Dom$ be a pre-Hilbert space with Hilbert space closure
$\Hilb=\overline{\Dom}$. Let $\pi$ be an
algebra homomorphism of the $*$-algebra $\A$
into the linear operators on $\Dom$,
$\pi:\, \A\ni a\mapsto \pi(a)$
such that $1\mapsto 1\restricted\Dom=\id_{\Dom}$.
Then $(\pi,\Dom)$ will be called a \Def{Hilbert space
representation} of $\A$.
If the algebra involution $*$ becomes represented by the
Hilbert space adjoint (the Hilbert space adjoint
of the operator $A$ will always be denoted by $A^\dagger$),
$\pi(a^*)=\pi(a)^\dagger_{|\Dom}$,
then $(\pi,\Dom)$ will be called a \Def{$*$-representation
of the $*$-algebra $\A$ in the Hilbert space $\Hilb$}
\cite{Schmud}.

For each $*$-representation $(\pi,\Dom)$
of a $*$-algebra $\A$ in a Hilbert space $\Hilb$ one defines its
\Def{adjoint representation} $(\pi^\dagger,\Dom^\dagger)$ with
\[
 \Dom^\dagger = \Dom(\pi^\dagger):=\bigcap_{a\in\A}
  \Dom\left(\pi(a)^\dagger\right) \quad\mbox{and}\quad
 \pi^\dagger(a) := {\pi(a^*)^\dagger}\restricted{\Dom^\dagger}
\]
which is a representation of $\A$ but not necessary a $*$-representation.
If a $*$-represent\-ation $\pi$ fulfills
$\pi=\pi^\dagger$ then it is called a \Def{self-adjoint representation}
\cite{Powers,Schmud}.

\begin{Satz}{Definition.}{Hilb of Obs}
The state space of $SO_q(3)$-symmetric quantum mechanics is
a separable Hilbert space $\Hilb$ with the following
additional structure:
\begin{Punkt}
\item $\Hilb$ carries a unitary continuous corepresentation
$(U,\Hilb)$ of the compact matrix quantum group $(\QGC,M)$.
\item
$\Hilb$ carries a $*$-representation $(\pi,\Dom)$
of the observable algebra $\Obs$ with $\Dom\densesubset\Hilb$.
\item
The scalars $X^2:=\scalar XX$ and $P^2:=\scalar PP$ in $\Obs$
become represented by observables: \\
{\rm $\pi(X^2),\pi(P^2)$ are essentially
self-adjoint operators in $\Hilb$.}
\item
The elements of $\Obs$ are represented by covariant operators:
{\rm\[
 U\, \left(\pi(a)\otimes 1_\QGC\right) \,U^* = \Delta_R(a)
\]
for each element $a\in\Obs$.}
\item
The $SO_q(3)$-symmetry manifests itself completely in the
observable algebra: \\
{\rm All elements $a\in\Obs$ in the angular momentum algebra,
i. e. $[a,P^2]=[a,X^2]=0$,
belong to the $*$-representation of $\QG'$ induced by $(\Hilb,U)$:
${\tilde a^U}_{|\Dom} = \pi(a)$.}
\end{Punkt}
If all these requirements are fulfilled, then $(\pi,\Dom)$ is called
a \Def{quantum mechanical representation of $\Obs$}.
\end{Satz}

\Absatz{Spin degrees of freedom.}{spin}
For the vector operators $\{X_i,P_i\}$ the covariance means certainly
$U (\pi(X_i)\otimes 1_\QGC) U^* = \pi(X_j)\otimes \M ji$ and
$U (\pi(P_i)\otimes 1_\QGC) U^* = \pi(P_j)\otimes \M ji$.
This implies that the operators $WS_i$ and $W^2$ can be identified
to belong to the representation of $\UUU$ on the observable algebra.
But (v) is not implied by (iv) which
only states that the $SO_q(3)$-covariance manifests itself on the
observable algebra by the angular momentum algebra
as orbital angular momentum. There could be hidden, i.~e.
internal, degrees of freedom also transforming under the symmetry perhaps
even under the quantum group $SU_{\sqrt q}$.
Let $(u,h)$ be some finite-dimensional
irreducible continuous corepresentation of $(\QGCU,m)$ ($m$ being the
matrix of the fundamental representation).
Let $\tilde\Hilb := \Hilb\otimes h$ and
define the unitary operator $\tilde U:=(\id\otimes \id\otimes \mult)\circ
 \sigma_{23} (U\otimes u)$.
Let the observable algebra $\tilde\Obs:=\Obs\otimes \UUU^U$ be extended
by the representation of $\UUU$ on $\Hilb\otimes h$.
Then also half-integer values for $l$ arise. In this way
spin degrees of freedom have to be introduced in the observable algebra
to characterize a quantum mechanical state completely.
However these possibility is from now on excluded by \refxy{Hilb of Obs} (v).

Because of \refxy{Hilb of Obs} (iv) the $\pi$-representatives of
the scalar operators $P^2$ and $X^2$ of $\Obs$ commute on $\Dom$
with the unitary corepresentation operator $U$
and hence also, because of (v), with operators
$\tilde\chi^U$ derived from $\tilde\chi\in\QG'$.
Then one may hope that the operators $\pi(X^2)$ (or $\pi(P^2)$
respectively), $\tilde W^U$ and $\tilde K^U$
can be simultanously diagonalized on $\Hilb$.
For such problems the following theorem will be used.

\begin{Satz}{Proposition:}{P714}
Let $\Dom\densesubset\Hilb$ be a common domain for the
symmetric Hilbert space operators $a$, $c$ and $c'$.
Let all three operators mutually commute on $\Dom$. Let
$a$ be essentially self-adjoint, and $\overline a \ge 1$.
Let $c$ and $c'$ be $a$-bounded (i.~e.
$\exists\lambda>0:\,\|c\phi\|\le\lambda(\|\phi\|+\|a\phi\|)$
for all $\phi\in\Dom$). \\
Then $c$ and $c'$ are essentially self-adjoint on $\Dom$.
Any other core for $\overline a$ is a core for $\overline c$
and $\overline {c'}$. All three operators
mutually commute strongly.
\end{Satz}

A core of a
closed Hilbert space operator $A$ with domain $\Dom_A$ is
a linear subspace $\Dom'\subset\Dom_A$ such that
$\overline{A\restricted{\Dom'}}=A$.
The notion of strong commuting of operators used here
is borrowed from \cite{Schmud} and
coincides with the notion of commuting in \cite{RS}.

\begin{Beweis}
That $\overline{c}$ and $\overline{c'}$ are self-adjoint on
any core for $a$ is a direct consequence
of \cite{RSA} with $N=a$ and $A=c,c'$.
That $c$ and $c'$ commute strongly is contents of \cite{SchmudA}
That $a$ and $c$ or $c'$ respectively commute strongly is contents
of \cite{SchmudB}.
\end{Beweis}

\begin{Satz}{Proposition:}{comm.ops}
In each quantum mechanical representation $(\pi,\Dom)$
the operators $\pi(X^2)$, $\tilde K^U$ and $\tilde W^U$
commute strongly and can be simultanously diagonalized.
This holds also for the operators $\pi(P^2)$,
$\tilde K^U$ and $\tilde W^U$.
\end{Satz}

\begin{Beweis}
The algebra $\QG'$ acts on $\QG$ by
$\tilde\rho\mapsto(\id\otimes\tilde\rho)\circ\Delta$
and the space $\QG$ decomposes in eigenspaces
$\QG(l,m):=\Span\{t^l_{n,m}:\> n=-l,-l\!+\!1,...,l\}$
of $\tilde K$ and $\tilde W$ with eigenvalues $q^{-m},c_l>0$.
One defines the linear functionals $\tildeE_{lm}\in\QG'$ by
\[
 {\tildeE_{lm}}{}_{\textstyle |\QG(l',m')} = \delta_{ll'}\,\delta_{mm'}.
\]
Then the operators $\tildeE_{lm}^U$
constructed according to \refxy{lm-states}
are the simultanous
spectral projectors of $\tilde K^U$ and $\tilde W^U$. They are
bounded operators on $\Hilb$ which commute with
$\pi(X^2)=\sum_i \pi(X_i) \pi(X_i^*)$
on $\Dom$. Then \refxy{P714} applies with
$c=\tildeE^U_{lm}{}\restricted\Dom$
and $a=1\restricted\Dom+\pi(X^2)$. It states that
$\pi(X^2)$, $\tilde K^U$ and $\tilde W^U$ commute strongly. Using the
symmetry $P^2\leftrightarrow X^2$ one proves the statement for $P^2$.
\end{Beweis}

{}From the above proof the definition of the simultanous
spectral projectors $\tildeE^U_{lm}$
of $\tilde K^U$ and $\tilde W^U$ derived from
the corepresentation operator $U$ will be used also in the sequel.

\begin{Satz}{Proposition:}{ESA-Rep}
In each quantum mechanical representation $(\pi,\Dom)$ the
representatives of the real generators $X^R_\alpha$ and
$P^R_\alpha$ of $\Obs$ are essentially self-adjoint.
The adjoint representation $(\pi^\dagger,\Dom^\dagger)$
is self-adjoint.
\end{Satz}

\begin{Beweis}
Proposition \refxy{P714} with $c=\pi(X^R_\alpha)$ and
$a=1\restricted\Dom+\pi(X^2)$ with $\lambda=\frac2{q+\frac1q}$ applies.
It states that $\overline{\pi(X^R_\alpha)}$ is
essentially self-adjoint for any core of $\pi(X^2)$.
The symmetry $X\leftrightarrow P$ proves the same for the momenta.
This implies the self-adjointness of $\pi^\dagger$ as proved
in \cite{SchmudC}.
\end{Beweis}

That means that every quantum mechanical representation can be uniquely
extended to a self-adjoint representation by simply taking the
adjoint representation. In all the following only the
self-adjoint quantum mechanical representations will be considered.

\Absatz{}{kernel}
In a quantum mechanical representation $(\pi,\Dom)$
the self-adjoint operators $\overline{\pi(X^2)}$ and
$\overline{\pi(P^2)}$ are positive, and their kernels are trivial.

\begin{Beweis}
Suppose $0\ne\ket\psi\in\Hilb$ is in the kernel of $\overline{\pi(X^2)}$.
Without restriction $\ket\psi= \tildeE_{lm}^U\ket\psi$.
Let $\ket f\in\Dom$. Then it is
$ 0=\linebreak[3]\bracket{\psi}{\pi(X^2)\pi(P^2)|f} = \linebreak
-\cequ^2
\bracket{\psi}{(q^{2l}-\pi(\mu^2))(q^{2l+2}-\pi(\mu^2))\pi(\mu^{-2})|f}$.
For all $l\ge 0$ this is in contradiction with $\mu^*{}^2\mu^2=q^{-6}$.
The positivity follows from $X^2=X_iX_i^*$.
\end{Beweis}

\Absatz{Spectral measures.}{spectal measure}
Since $\overline{\pi(X^2)}$ and $\overline{\pi(P^2)}$ are self-adjoint
in a quantum mechanical representation $(\pi,\Dom)$ the
spectral theorem (e.~g. \cite{RSB}) applies and one can
introduce the projection-valued measures $\{\specm X\Omega\}$ and
$\{\specm P\Omega\}$ of $\overline{\pi(X^2)}$ and
$\overline{\pi(P^2)}$ which fulfill
\begin{eqnarray*}
 &\overline{\pi(X^2)} = \int_{x\in\RR_+} x^2\, {\bf d}\specm Xx\quad
 &\mbox{with}\quad
 \specm X\Omega=\int_\Omega {\bf d}\specm Xx,\nonumber \\
 &\overline{\pi(P^2)} = \int_{p\in\RR_+} p^2\, {\bf d}\specm Pp\quad
 &\mbox{with}\quad
 \specm P\Omega=\int_\Omega {\bf d}\specm Pp.
\end{eqnarray*}
Because of \refxy{comm.ops} the spectral projectors of either of
the sets $\{\pi(X^2),\tilde W^U,\tilde K^U\}$ or
$\{\pi(P^2),\tilde W^U,\tilde K^U\}$ commute and the spaces
\[
 \Hilb^X_{\Omega,l,m}:=\specm X\Omega\,\tildeE_{lm}^U\Hilb
  \quad\mbox{and}\quad
 \Hilb^P_{\Omega,l,m}:=\specm P\Omega\,\tildeE_{lm}^U\Hilb
\]
(to be thought of as approximate simultanous eigenspaces)
span dense subspaces
\[
 \hat\Dome^X := \sum_{0<a\le b<\infty\atop|m|\le l\in\NN_0}
      \Hilb^X_{[a,b],l,m}
\qquad\mbox{and}\qquad
 \hat\Dome^P := \sum_{0<a\le b<\infty\atop|m|\le l\in\NN_0}
      \Hilb^P_{[a,b],l,m}
\]
in the Hilbert space $\Hilb$.

\Absatz{}{D-Hut}
Let $(\pi,\Hilb)$ be a self-adjoint quantum mechanical representation of
$\Obs$ on $\Hilb=\overline\Dom$. Then the following statements hold:
\begin{Punkt}
\item $\hat\Dome^X\subset\Dom$ is invariant under $\pi$.
I.~e. the restricted representation
$(\pi\restricted{\hat\Dome^X},\hat\Dome^X)$
is a $*$-representation of $\Obs$.
\item $\hat\Dom:=\hat\Dome^X + \hat\Dome^P$ is invariant under $\pi$.
The $*$-representation
$(\hat\pi,\hat\Dom):=(\pi\restricted{\hat\Dom},\hat\Dom)$ determines
the representation $\pi$ completely with
$(\pi,\Dom)=(\hat\pi^\dagger,\hat\Dom^\dagger)$.
\end{Punkt}

\begin{Beweis}
Since $\hat\Dome^X$ is a core of $\overline{\pi(X^2)}$, \refxy{P714}
can be applied with $c=\pi(X_\alpha^R)$, and one concludes
that $\hat\Dome^X\subset\Dom(\pi(X_\alpha^R)^\dagger)$ is a core for
$\overline{\pi(X_\alpha^R)}$ and that
$\hat\Dome^X$ is invariant, since the eigenvalues of the operators
$\tilde K^U$ and $\tilde W^U$ are changed by at most one unit.
The action of the operators $\pi(\mu^{\pm 1} L_i)^\dagger$ and
$\pi(\mu^{\pm 1} W)^\dagger$ can be derived from \refxy{lm-states} and
$\pi(\mu W)\Hilb_{[a,b],l,m} = \linebreak[3]
\Hilb_{[q^{\mp 1} a,q^{\mp 1} b],l,m}$ proving the invariance of
$\hat\Dome^X$ under
these operators. Finally, observing $\pi(X^2)^\dagger\hat\Dome^X=\Dome^X$,
the same conclusions can be drawn for $\pi(P_i)$ when \refxy{reconstr}
is taken into account. This proves $\hat\Dome^X\subset\Dom^\dagger=\Dom$.
The symmetry $P\leftrightarrow X$ assures the same conclusions
for $\hat\Dome^P$.
Thus $\hat\Dom\subset\Dom$ is an invariant subset of the domain of $\pi$.
Since the real generators $\{X_\alpha^R,P_\alpha^R\}$
of $\Obs$ are represented by essentially self-adjoint operators on
$\hat\Dom$ the last assertion follows \cite{SchmudD}.
\end{Beweis}

\begin{Satz}{Proposition:}{l>0}
Let the conditions and definitions be as in \refxy{D-Hut}.
Let $\hat\Dome^X_{lm}:=\tildeE_{lm}^U\hat\Dome^X
\densesubset\tildeE_{lm}^U\Hilb$.
Then the restriction
$\pi(P^2)\restricted{\hat\Dome^X_{lm}}$ is an essentially self-adjoint
operator in the Hilbert space $\tildeE_{lm}^U\Hilb$.
Its spectral measure is given by
\begin{eqnarray*} \label{specmeasure}
 {\bf d}\specm Pp\tildeE^U_{lm} &=&
  \sum_{k,k'\in\ZZ} q^{6(k+k')} \;      N_l^{-2}\, F_l(q^{\frac12+l+2k})\,
             \overline{F_l(q^{\frac12+l+2k'})} \>\times\nonumber\\
 &&\qquad\quad\times\>
    \pi({\mu^*}^{2k}) {\bf d}\specm X{\textstyle\left[{q^{\frac12+l}
\cequ}p^{-1}\right]}
                 \tildeE^U_{lm} \pi(\mu^{2k'})
\end{eqnarray*}
with the $q$-deformed Bessel functions
\[
  F_l(y) = y^l \sum_{k=0}^\infty y^{2k} (-1)^k
   \frac{ q^{-2k (k-1)} q^{-k(2l+5)} }{ (q^{-4};q^{-2(2l+3)})_k
      (q^{-4};q^{-4})_k},
\]
the definitions
$(a;p)_k := \prod_{j=0}^{k-1} (1-ap^j)$ and
$N_l := q^{-l(l+\frac12)}\frac{ (q^{-4};q^{-2(2l+3)})_\infty }
{ (q^{-4};q^{-4})_\infty }$ and the constant
$\cequ = \frac{q^2}{(1+q^3)(1-q)}$ introduced in \refxy{reconstr}.
\end{Satz}

The function $F_l$ can also be expressed with the help of the
$q$-hypergeometric functions discussed in \cite{Ko}:
$F_l(y)=y^l\hyp{\textstyle
 \left(0;q^{-2(3+2l)};q^{-4},y^2 q^{-2l-5} \right)}$.
\cite{KoA} states then
\[
 \delta_{nm} = N_l^2 q^{3(m+n)} \sum_{k\in\ZZ} q^{6k} \,
   F_l(q^{\frac12+l+2(k+m)})\, F_l(q^{\frac12+l+2(k+n)})
\]
when $l\ge -1$.
Introducing the operators
$ (K_l f)(y) := q^{-l}f(qy) - q^l f(q^{-1} y), $ and
$ \Delta_l f (y) := \frac{\cequ^2}{y^2} K_{-l-1} K_l f(y)$
the functions $F_l$ fulfill
the following difference equations ($l\in\NN_0$):
\begin{eqnarray*}
 & \frac1y\, (K_l F_l)(y) = {\textstyle \frac{q^{-2l-3}}{1-q^{-2(2l+3)}}}\,
   F_{l+1}(y), &\nonumber\\
 & \cequ^{-2}\Delta F_l(y) = \frac1{y^2} K_{-l-1} K_l F_l (y) = F_l(y)&
\end{eqnarray*}
which should be compared with \refxy{reconstr}.

\begin{Beweis}
To prove the essential self-adjointness of
$\pi(P^2)\restricted{\hat\Dome^X_l}$ its defect indices are
investigated. Let $\Psi$ be a $\pm i$-eigenvector of
$\pi(P^2)\restricted{\hat\Dome^X_l}^\dagger$.
Then $ 0 = (\pi(P^2)\pm i) \, \specm X{\Omega} \ket\Psi$.
Thus, using the formulas \refxy{reconstr},
almost everywhere
$0=(-\Delta_l \pm i) {\rm d}\Psi_y$ with ${\rm d}\Psi_y := y^{-\frac32} {\bf
d}\specm Xy\Psi$
This difference equation gives a dependence between ${\rm d}\Psi_{q^{2n}y}$
for $y\in [1,q^2)$, but does not correlate different $y\in [1,q^2)$.
Hence for each $y\in [1,q^2)$ it leads to a recursion relations which
has a two-parametric solution space. The general solution is thus
${\rm d}\Psi_{q^{2n}y} = \left(A_y F_l(\pm i\frac1c y q^{2n})
  + B_y F_{-l-1}(\pm i\frac1c y q^{2n})\right) {\rm d}\Psi_y $
with measurable $A_y$ and $B_y$ for $y\in [1,q^2)$.
However there is still the norm squared of $\Psi$ given by
$\bracket\Psi\Psi=
 \int_{y\in[1,q^2)} \bracket\Psi{y^{-3} {\bf d}\specm Xy|\Psi}
 \sum_{n\in\ZZ} q^{6k} \left|A_y F_l(\pm \sqrt i\frac1c y q^{2n})
  + B_y F_{-l-1}(\pm \sqrt i\frac1c y q^{2n})\right|^2 $ .
Hence the sums must be finite almost everywhere.
For $l\ge 1$ it is now easy to see that $B_y$ has to vanish almost
everywhere. Otherwise the partial sums
above would diverge for $n\to\-\infty$.
On the other hand one sees that the $n\to-\infty$-limit gives no conditions
for $A_y$.
But it will be shown that $|F_l(\pm i y)|\to\infty$ for $y\to\infty$
which proves that also $A_y$ has to vanish almost everywhere
thus implying that $\ket\Psi=0$.
A strongly related analysis is in \cite{HWA}.
$f(r) := i^{\frac l2} q^{-l(l+\frac12)}F_l(\frac{\pm i}c y) = r^l
 \sum_k (\pm i)^k q^{-2k(k+1)} r^{2k} G_{k,1} $
with
$G_{k,m} := \linebreak[4]
\left((q^{-4};q^{-4(l+m)+\frac12})_{(k-m+1)}\,
                       (q^{-4};q^{-4m+4})_{(k-m+1)}\right)^{-1}$ and
$r=\frac{q^{-l(l+\frac12)}}c y$. To investigate the limit $r\to\infty$
one chooses $r=q^{2(t+\alpha)}$ with $t\in 4\NN_0$, $\alpha\in\RR$.
Substituting into $f$ this gives $f(r)=\Phi_1 + \Phi_2$ with
\linebreak[3]
$\Phi_2=q^{2l(t+\alpha)} \sum_{k=\frac t2+1}^\infty G_{k,1}
q^{-2k(k+1-2t-2\alpha)}(\pm i)^k \linebreak[4]
=:G_{\frac t2,1} q^{2l(\alpha+t) + 2(\alpha+t-\frac12)^2}
\psi^{(i)}_t(\alpha)$.
Using the limit $G_{t+k,\frac t2+1}\to 1$ for $t\to\infty$ one has
$\lim_{t\to\infty} \psi^{(i)}_t(\alpha) =
\psi^{(i)}(\alpha)\linebreak[2]:=\sum_{k=-\infty}^\infty
 (\pm i)^k q^{-2(k+\frac12-\alpha)^2}$.
For large $t$, the absolute terms of the individual terms of the
alternating sums of real and imaginary
part of $\Phi_1$ grow monotonously with $k$ and the sum can
be estimated using the last term:
$|\Phi_1|< 2G_{\frac t2,1} q^{\frac32 t^2 - t(q-2\alpha-l) +2\alpha l}$.
Then one arrives at
\[
 \lim_{t\to\infty\atop t\in 4\NN_0}
 \frac{|f(q^{2(t+\alpha)})|}{\phi(t,\alpha)}=\psi^{(i)}(\alpha)
\]
with the monotonously growing function
$\phi(t,\alpha)=q^{2t^2-2t(1-2\alpha-l)+2\alpha l +2(\frac12-\alpha)^2}$.
The function $\psi^{(i)}$ is continuous and periodic with period $4$.
Let $\psi'$ be its real part, then
$\psi^{(i)}(\alpha)=\psi'(\alpha)\pm i\psi'(\alpha-1)$.
In \cite{HW} it is proved that $\psi'(\alpha)=0$ implies
$\alpha=-\frac12+2\ZZ$. Thus $|\psi^{(i)}|$ is strictly positive.
Hence the sum above would diverge for $n\to\infty$. Therefore
$A_y$ has to vanish almost everywhere. \\
Doing the same analysis as in \cite{HW} or using the results
of \cite{Ko} cited above gives the explicit spectral representation.
\end{Beweis}

Since the image of ${\bf d}\specm Pp\tildeE^U_{lm}$ belongs to the domain
$\hat\Dom\subset\Dom$ the operators $\pi(P_i)$ can be
applied. Therefore the term
$\tildeE^U_{00} \frac1p\pi(P_i)\tildeE^U_{1m}\>
 {\bf d}\specm Pp\>
 \tildeE^U_{1m'} \frac1p\pi(P_j)\tildeE^U_{00}$ leads to the still missing
spectral representation of $\pi(P^2)$ for $l=0$, and one gets the
following:

\begin{Satz}{Corollary:}{Cor}
The spaces $\hat\Dome^P$ and hence
$\hat\Dom$ are uniquely determined by $\hat\Dome^X$, since the equation
\begin{eqnarray}
 {\bf d}\specm Pp\tildeE^U_{lm} &=&
  \sum_{k,k'\in\ZZ} q^{6(k+k')} \;      N_l^{-2}\, F_l(q^{\frac12+l+2k})\,
             \overline{F_l(q^{\frac12+l+2k'})} \>\times\nonumber\\
 &&\qquad\quad\times\>
    \pi({\mu^*}^{2k}) {\bf d}\specm X{\textstyle\left[{q^{\frac12+l}
\cequ}p^{-1}\right]}
                 \tildeE^U_{lm} \pi(\mu^{2k'})
\end{eqnarray}
is valid for all $l\in\NN_0$.
\end{Satz}

Thus to classify all the self-adjoint quantum mechanical representations
it is sufficient to construct the spaces $\hat\Dome^X$ representing $\Obs$.

\Absatz{}{l=0}
Each nontrivial self-adjoint representation of $\Obs$ contains
some vector $0\ne\ket\Psi\in\Hilb_{[1,q),0,0}$.

\begin{Beweis}
There exists $0\ne\ket\Phi\in\Hilb_{[a,b],l,l}$.
One observes that the reducible multiplet
$\pi(X_3)\ket\Phi\in\Hilb_{[a,b],l+1,l-1}+\Hilb_{[a,b],l-1,l-1}$
with $\Phi\in\Hilb_{[a,b],l,l}$ because of
$ 0=\tildeE^U_{ll}\, \pi(\scalar XL)\ket{\Phi} =
 q^{-1/2}\,\tildeE^U_{ll}\, \pi(L_1 X_3)\ket{\Phi} +
 \tildeE^U_{ll}\, \pi(L_2 X_2)\ket{\Phi} = $
$ -q^{-\frac12}\,[l+1]_q\,\tildeE^U_{ll}\, \pi(X_2)\ket{\Phi}$.
Then one compares the vectors
constructed as $\pi(X_3)\ket{r,l,l}$ and $\pi(L_3 X_2)\ket{r,l,l}$.
\[
 \frac{ \left\|\pi(X_3)\ket{\Phi}\right\|^2
        \left\|\pi(L_3 X_2)\ket{\Phi}\right\|^2}
  { \left|\bracket{\Phi}{\pi(L_3 X_2)^\dagger\,\pi(X_3)|\Phi}\right|^2 }
 = q^l\,[2l+1]_{\sqrt q}\,{\textstyle
  \left(1 + q^{-2l}\frac{[2l]_{\sqrt q}}{[2]_{\sqrt q}}\right)} \ge 1.
\]
One concludes that they are linearly independent for $l>0$, and
$\Hilb^X_{[a,b],l-1,l-1}$ is nontrivial if $l>0$. Since the
operator
$u:=q^{\frac32}\, \overline{(\tilde W^U)^{-1}\, \pi(\mu W)}$
provides an isometry between the spaces
$\Hilb^X_{[q^n a,q^n b]}$ for $n\in\ZZ$ the lemma is proved.
\end{Beweis}

\Absatz{Irreducible representations.}{Psi-Darst}
Defining $\hat\Dome_\Psi:=\pi(\Obs)\ket\psi$,
$\Hilb_\Psi:=\overline{\hat\Dome_\Psi}$, $\pi_\Psi:=\pi\restricted\Hilb_\Psi$
one gets now a self-adjoint quantum mechanical representation of $\Obs$ on
the Hilbert space $\Hilb_\Psi$. Then its complement is also
a self-adjoint quantum mechanical representation \cite{SchmudE}.
One observes also at once that this
subrepresentation is irreducible if, and only if,
$\ket\Psi\in\specm X{[1,q)}$ is an eigenvector of $\pi(X^2)$.

Let us suppose now that $(\pi,\Dom)$ is irreducible. Take $\ket\Psi$
as before and suppose that it is
normalized, $\bracket\Psi\Psi=1$, and obeys $\pi(X^2)\ket\Psi=x_0^2\ket\Psi$
with $x_0\in[1,q)$. This irreducible representation will be called
from now on $(\pi_{x_0},\Dom_{x_0})$.
Then one defines the following orthonormal basis of $\hat\Dome^X_{x_0}$
\[
 \left\{ \ket{x_n,l,m}\biggl| n\in\ZZ,\> l,|m|\in\NN_0,\>|m|\le l\right\}
\]
of simultanous eigenvectors of the
commuting observable $\pi_{x_0}(X^2)$ with eigenvalues
$x_n^2 = x_0^2 q^{2n}$ and $\pi_{x_0}(W^2)$ and $\pi_{x_0}(L_2W)$
by
\begin{eqnarray*}
 && \ket{x_n,0,0} =
  {u^\dagger}^n\ket\Psi:=\pi(q^{-\frac32}\mu^{-1} W)^n\ket\Psi \nonumber\\
 && \ket{x_n,l,l}:=
  \sqrt{\frac{[2]_{\sqrt q}}{q^n x_0}
  \frac{[2l+3]_{\sqrt q}}{[2l+2]_{\sqrt q}} } \pi_{x_0}(X_1)
  \ket{x_n,l-1,l-1} \\
 && \ket{x_n,l,m} := q^{-\frac12} \alpha_{l,m-1}^{-1} \ket{x_n,l,m+1}.
 \nonumber
\end{eqnarray*}
In this irreducible representation also $\pi_{x_0}(P^2)$ has exactly
one eigenvalue in $[1,q^2)$.
To complement $\hat\Dome^X_{x_0}$ one has still to determine
all the eigenvectors of $\pi_{x_0}(P^2)$. Defining
\[
 p_0 := \frac {\sqrt q\cequ}{x_0}
\]
the simultanous eigenvectors $\ket{p_n,l,m}$ of $\pi_{x_0}(P^2)$
with eigenvalues $p_0^2 q^{2n}$ and $\pi_{x_0}(W^2)$ and
$\pi_{x_0}(WL_2)$ in the same manner which span $\hat\Dome^P_{x_0}$
one has still only to give the matrix elements between these both
bases. But this can be read off from \refxy{l>0}:
\[
 \bracket{x_n,l,m}{p_{n'},l',m'} = \delta_{ll'}\,\delta_{mm'}\,
            \frac1{N_l} F_l(q^{n+n'+l+\frac12}) \delta^{(2)}_{n+n'+l}
\]
with $2\delta^{(2)}_a := (-1)^a + 1$.

Now the results \refxy{comm.ops}, \refxy{ESA-Rep} and \refxy{Psi-Darst}
can be summarized as a theorem:

\begin{Satz}{Theorem:}{TT}
Each quantum mechanical representation of $\Obs$ can be
uniquely extended to a
self-adjoint quantum mechanical representation by taking the
adjoint representation.
Each self-adjoint quantum mechanical representation $(\pi,\Dom)$
is a direct integral sum of the
irreducible quantum mechanical representations $(\pi_{x_0},\Dom_{x_0})$
parametrized by $x_0\in [1,q)$, and it holds upto unitary equivalence
\[
 \Dom \sim \int^\oplus_{x_0\in [1,q)} \Dom_{x_0}
        \otimes {\bf d}\specm X{x_0}\tildeE^U_{00} \Hilb, \qquad
 \pi(a) \sim  \int^\oplus_{x_0\in [1,q)} \pi_{x_0}(a)
        \otimes\id_{{\bf d}\specm X{x_0}\tildeE^U_{00} \Hilb}
\]
for $a\in\Obs$.
The operator sets $\{\pi(X^2),\pi(W^2),\pi(L_2)\}$ and
$\{\pi(P^2),\pi(W^2),\pi(L_2)\}$ respectively are complete
sets of commuting observables.
\end{Satz}

Thus after having characterized a quantum mechanical representation
by very natural requirements on the observability of certain operators
and the covariance under the quantum group all representations are
found. In opposite to the usual quantum mechanical situation where
von Neumann's arguments show the uniqueness of the Schr\"odinger
representation for the Heisenberg algebra \cite{JvN},
in the $q$-deformed case a one-parametric set of inequivalent
irreducible quantum mechanical representations arise which however have the
property that eigenvectors of position and momentum exist.

These results can in the same way be obtained also for
$SO_q(N)$-symmetric quantum mechanics. The special case
$N=1$ is treated in \cite{HSSWW} where the fact that
$\pi(P^2)$ is not essentially self-adjoint on $\Dome^X$ leads
to a lot of ambiguities.

\section{Application: Potential Problems}\label{PotProblems}
In analogy to usual quantum mechanics one can now consider
properties of operators of the form $H=\pi(P^2) + V$ and solve
the corresponding eigenvalue equations.

\Absatz{}{function norms}
Let $(\pi_{x_0},\Dom_{x_0})$ be the irreducible
self-adjoint quantum mechanical representation of $\Obs$ with
$\pi_{x_0}(X^2)$-eigenvalues $x_0^2 q^{2n} = x_n^2$.

For sequences $f: \ZZ\to\CC$ the mappings
$f\mapsto ||f||_\infty$, $f\mapsto ||f||_2$, $f\mapsto ||f||_\infty$
are defined by $||f||_\infty:=\sup_{n\in\ZZ} |f_n|$,
$||f||_1:=\sum_{n\in\ZZ} q^{3n} |f_n|$, $||f||_2:=\sqrt{||f^2||_1}$.
These sequences can be understood as multiplication operators in
the $x_0$-representation with the definition
$\pi_{x_0}(f)\ket{x_n,l,m}:=f_n\ket{x_n,l,m}$
with domain to be specified.
Certainly the sequences $f$ with $||f||_\infty<\infty$ act as
bounded operators. Then in the spirit of
\cite{RSC} one can prove the following result.

\begin{Satz}{Proposition:}{X.15}
Let $(\pi_{x_0},\Dom_{x_0})$ be the irreducible quantum mechanical
representation of $\Obs$ as before.
Let $v$ and $w$ be sequences $\ZZ\to\CC$,
$||v||_\infty<\infty$ and $||w||_2<\infty$ acting
as multiplication operators in the $x_0$-representation.
In addition let $b$ be a bounded operator on $\Hilb_{x_0}$
which commutes with the corepresentation operator $U$
such that $b\pi_{x_0}(v+w)$ is a symmetric operator on
$\Dom_{x_0}$ (which
commutes then strongly with the operator $\tilde K^U$ and $\tilde W^U$).
Then $H := \pi_{x_0}(P^2) + b\pi_{x_0}(v+w)$ is essentially self-adjoint
on $\Dom_{x_0}$.
\end{Satz}

\begin{Beweis}
Let $\Hilb:=\overline\Dom_{x_0}$.
That $H$ is essentially self-adjoint will be proven separately for
the subspaces $\tildeE^U_{lm}\Hilb$.
The vectors in $\ket\phi\in\tildeE^U_{lm}\Hilb$ define
sequences $\phi_n:=q^{-\frac32 n}\bracket{x_n,l,m}\phi$
and $\hat\phi_n:=q^{-\frac32 n}\bracket{p_n,l,m}\phi$
with $||\phi||_2=||\hat\phi||_2<\infty$. Then
\[\label{IX.28a}
 ||\phi||_\infty\le\sum_{k\in\ZZ}
 \sup_{n+k}\left|q^{\frac32(n+k)}\bracket{x_{n+k},l,m}{p_0,l,m}\right|
 \>q^{3k} |\hat\phi_k| \le d_l ||\hat\phi||_1
\]
with the positive constant $d_l$. \\
On the other side the sequence $\Pi:\, n\mapsto \frac1{p_n^2+1}$
has finite norm $||\Pi||_2$. Let
$\ket\phi\in\tildeE^U_{lm}\Dom_{\pi_{x_0}(P^2)^\dagger}$ be in the
domain where $\pi_{x_0}(P^2)^\dagger$ restricted to
$\tildeE^U_{lm}\Hilb$ is self-adjoint.
Then $(1+\pi_{x_0}(P^2))\ket\phi\in\Hilb$ lets define the sequence
$\Pi^{-1} \hat\phi$. With Schwarz's inequality
$||\hat\phi||_1\le ||\Pi||_2\,||\Pi^{-1} \hat\phi||_2 \le
||\Pi||_2\left( ||\pi_{x_0}(P^2)\ket\phi|| + ||\ket\phi||\right)$.
Applying these inequalities to the vectors $\mu^r\ket\phi$ results in
\[\label{IX.28b}
 ||\hat\phi||_1 \le q^{\frac12 n} ||\Pi||_2 \>||\ket\phi|| +
 q^{-\frac32 n}||\Pi||_2 \>||\pi_{x_0}(P^2)\ket\phi||.
\]
This formula implies that for each
$\phi\in\Hilb_l$ and each $e>0$
one can find $g_l>0$ (depending on $l$ and $e$ but not on the
vector $\ket\phi$) such that
\[\label{IX.28c}
 ||\phi||_\infty \le e ||\pi_{x_0}(P^2)\ket\phi|| + g_l ||\ket\phi||.
\] \\
Let the operator norm of $b$ be $||b||$.
Then one proves straight forwardly that choosing
$e:=\tau ||w||_2^{-1}\,||b||^{-1}$
and $\gamma:=g_l||w||_2 ||b|| + ||v||_\infty ||b||$
with $0<\tau<1$ leads to the inequality
\[\label{IX.28d}
 ||b(v+w)\ket\phi|| \le \tau||\pi_{x_0}(P^2)\ket\phi|| + \gamma ||\ket\phi||.
\]
Then the Kato-Rellich Theorem \cite{RSD} proves that
$H\restricted{\tildeE^U_{lm}\Hilb}$
is essentially self-adjoint on any core of
$\pi_{x_0}(P^2)\restricted{\tildeE^U_{lm}\Hilb}^\dagger$.
Thus $H$ is essentially self-adjoint.
\end{Beweis}

\Absatz{Example: The $SO_q(3)$-symmetric Coulomb problem.}{q-H-Atom}
Let $(\pi_{x_0},\Dom_{x_0})$ be as before.
Identify $a\in\Obs$ with $\pi_{x_0}(a)$.
Define the bounded operator $b$ by
$b:=-\mu-q^{-1}\mu^*$ where the bounded linear operator $\mu$ is
defined by
$\mu\ket{x_n,l,m}=q^{-\frac32} \ket{x_{n-1},l,m}$
(i.~e. this $\mu$ represents $\mu\in\DDR$).
The operator $\frac{1}{R}=\pi_{x_0}(v+w)$
is defined by the sequences $v$ and $w$ acting in the
$X$-representation and $(v+w)_n := \frac{1}{x_n}$ and $v_m=0$ for $m<0$.
(The such defined operator is uniquely defined by the
equation $(\frac1R)^2 \,X^2=1$ and its positivity.)
Then
\[
 H := P^2 + b\,\frac1R = P^2-(\mu\frac1R+\frac1R\mu^*)
\]
is essentially self-adjoint on $\Dom_{x_0}$. (It is obvious that
the potential $-\mu\frac1R-\frac1R\mu^*$ is
a $q$-deformation of the usual Coulomb potential.)

Like in the undeformed case one can find the Lenz-vector
consisting of the real triplet $\{A_i\}$ with
\[
 A_i:= -iq^2 \cequ\,[P_i,\tilde W^U] +
\frac{X_i}R
\]
which commutes with the Hamiltonian, $[H,A_i]=0$, and
closes on constant energy spaces together with the angular momentum algebra
in a $q$-deformed $SO(4)$-angular momentum algebra:
\begin{eqnarray*}
&& \cross AAk = -L_k\, W\, H,\nonumber\\
&& \scalar AA = \frac{q-1+\frac 1q}{q-2+\frac1q} W^2\, H
     - \frac1{q-2+\frac1q} H + 1,\nonumber\\
&& \scalar AL = \scalar LA = 0,\\
&& \cross LAk + \cross ALk = W A_k + A_k W,
 \nonumber\\
&& \cross LAk - \cross ALk = \frac{[2]_q}{(\sqrt q - \frac 1{\sqrt q})^2}
    [W,A_k].\nonumber
\end{eqnarray*}
The negative spectrum is given by
\begin{eqnarray*}
&\biggl\{
 E_n\biggl|\; n\in\NN\>, \;
 \exists {m\in\ZZ}:\, q^2 (q^{-n}\!-\!q^{n})\, E_n = x_0 q^m
  \biggr\} &\nonumber\\
&\mbox{with}&\\
& E_n = - {{\textstyle \left(\frac{1+q^{-3}}{1+q}\right)^2}}\,
 {\displaystyle {1\over [n]_q^2}} &\nonumber
\end{eqnarray*}
where the angular momentum is restricted like in the undeformed case
by $l+1\le n$.
Negative energies arise only for special representations
$(\pi_{x_0},\Dom_{x_0})$. This comes
from the fact that the corresponding wave functions are of the form \linebreak
$x^{n-1} \sum_k \frac{q^{-k(k+1)}}{(q^{-2};q^{-2})_k}
 \left( -q^2 (q^n\!-\!q^{-n})E_n \,x\right)^k$ which are only
normalizable for special values of $x_0$.

In this paper a complete framework for $SO_q(3)$-symmetric
one-particle quantum mechanics is provided and the representations
studied. The observable algebra is a deformation of the
usual Heisenberg algebra of coordinates and momenta
covariant under the Lie group $SO(3)$ which is obtained by
canonical quantization. However if quantization also changed the
symmetry as a second order effect but conserved the corepresentation
properties then the canonical quantization would
have to be modified. Considering $q=1+O(\hbar^2)$ the above found
algebra would be nothing else than a noncanonical $\hbar$-deformation
covariant under the $\hbar^2$-quantized group. \\[10pt]

The author thanks Bernhard Drabant, Moritz Fichtm\"uller,
Silke Grosholz, Hern\'an Ocampo, Joachim Seifert and Julius Wess
for their interest and numerous discussions.

\end{document}